\documentclass[final,5p,twocolumn,12pt]{elsarticle}

\usepackage{graphicx}
\usepackage{amssymb}
\usepackage{hyperref}
\biboptions{square,comma,sort&compress}

\journal{Journal of Crystal Growth}
\def\SFO{SrFeO\ensuremath{_{3-\delta}}}
\def\TN{\ensuremath{T_N}}

\begin{document}

\begin{frontmatter}

\title{Floating Zone Growth of Large Single Crystals of \SFO}

\author[FKF,UBC]{D.C. Peets\corref{1}}
\ead{dpeets@fkf.mpg.de}
\author[FKF]{Jung-hwa Kim}
\author[HMI]{M. Reehuis}
\author[UBC]{P. Dosanjh}
\author[FKF]{B. Keimer}

\address[FKF]{Max-Planck-Institut f\"ur Festk\"orperforschung, Heisenbergstr.\ 1, D-70569 Stuttgart, Germany}
\address[UBC]{Dept. of Physics \&\ Astronomy, University of British Columbia, 6224 Agricultural Rd., Vancouver, BC V6T~1Z4, Canada}
\address[HMI]{Helmholtz-Zentrum Berlin f\"ur Materialien und Energie, D-14109 Berlin, Germany}
\cortext[1]{Corresponding author}

\begin{abstract}

The Fe$^{4+}$-containing cubic perovskite phase \SFO\ is of interest both for its high-temperature, oxygen-conducting properties as a solid oxide fuel cell component, and at low temperatures, where it exhibits a plethora of helical magnetic phases and a candidate Skyrmion lattice.  However, a sequence of structural phase transitions encountered on cooling to room temperature has limited the size of single crystals.  We report the floating-zone growth and oxygen-annealing of multiple-cubic-centimetre-sized single crystals of \SFO, suitable for inelastic neutron scattering and other measurement techniques requiring large sample volumes.
\end{abstract}

\end{frontmatter}

\section{\label{sec:Intro}Introduction}

Pure and cation-doped versions of the cubic perovskite \SFO\ have been investigated for use as mixed conductors in solid oxide fuel cells~\cite{Huang2001,Ralph2001,Wincewicz2004} --- the material is not only electrically conducting, but is also a good conductor of oxygen ions at elevated temperatures, via oxygen vacancies.  At lower temperatures, closely-related perovskite-based manganates have been investigated extensively over the past two decades, primarily for exhibiting `colossal magnetoresistance'~\cite{Salamon2001,Tokura2006}.  The crystal structure of LaMnO$_3$, the progenitor of one of the most extensively studied manganate families, is that of \SFO\ but with tilting of the octahedra, and both materials have, in principle, the same high-spin $3d^4$ electron configuration --- \SFO\ is one of the few stable compounds in which the rare Fe$^{4+}$ state is realized~\cite{Gallagher1964,Fournes1987}.  Despite these similarities, however, these materials exhibit completely different electronic behaviour. While LaMnO$_3$ is a Mott insulator with orbital order and commensurate, collinear antiferromagnetism, stoichiometric SrFeO$_3$ is metallic~\cite{MacChesney1965} and orbitally degenerate at all temperatures, with significant charge transfer to the iron atoms from adjacent oxygens~\cite{Oda1977b,Bocquet1992,Dann1994,Mostovoy2005,Haas2009}.  It exhibits a transition to incommensurate, helical magnetic order at $\TN = 134$~K~\cite{Takeda1972}, and at least five distinct helical magnetic phases have been observed as a function of temperature and field at just this one oxygen content, including a candidate Skyrmion lattice phase~\cite{Ishiwata2011}.  The full evolution of the magnetic phase diagram with oxygen doping is not known, and the material remains of interest to the physics community.  One row below iron in the periodic table, SrRuO$_3$ is ferromagnetic~\cite{Callaghan1966}, while closely-related Sr$_2$RuO$_4$ is thought to be a rare triplet superconductor~\cite{Mackenzie2003}, and Sr$_3$Ru$_2$O$_7$ exhibits strange quantum critical points with quantum nematic fluid phases~\cite{Grigera2001,Borzi2007}.  Given the diversity of novel physics seen in the closely-related ruthenates and manganates, it is strongly desirable that the \SFO\ structural and magnetic phase diagrams be fully understood as a function of oxygen doping.

In principle, growth of single-crystalline \SFO\ should be straightforward by the floating-zone technique --- the compound melts congruently in air at 1435$^\circ$C~\cite{Maljuk2003,Fossdal2004}, and evaporation of the component oxides is not a significant problem at this temperature.  However, studies of \SFO\ have been hampered by a plethora of first-order structural phase transitions and phase separation regions in its oxygen phase diagram~\cite{Takeda1986,Mizusaki1992}.  Changes to the oxygen content on cooling from growth or sintering temperature to room temperature typically encounter several of these, so single crystals reach room temperature severely cracked, typically into pieces of a few cubic millimetres~\cite{Ishikawa1998,Maljuk2003}, with the largest reported being $\sim7\times7\times7$~mm$^3$~\cite{Maljuk2006}.  The dominant oxygen diffusion mechanism through such a crystal is via microcracks rather than oxygen vacancies, while the same cracks complicate the interpretation of transport measurements.  Detailed studies of the magnetic structure are typically performed by neutron diffraction and inelastic neutron scattering, the latter of which requires crystals, or mosaics of crystals, several cubic centimetres in size.  Minimizing the cracking would be very helpful for studies of the low-temperature transport properties, while eliminating it entirely would enable studies of the material's intrinsic oxygen diffusion properties.  Neutron scattering studies, the goal of our work, would be significantly more convenient if the crystal held together, but are otherwise largely insensitive to minor cracking.  Finally, while detailed studies of oxygen dynamics in the material have been performed~\cite{Mizusaki1992}, this knowledge has not yet been employed in enabling comprehensive doping-dependent studies of the low-temperature properties.  In this paper, we describe a technique for maximizing the size of image furnace-grown single crystals suitable for neutron scattering, and demonstrate their annealing and characterization.  

\section{\label{sec:expt}Crystal Growth}

\SFO\ powder was calcined from well-mixed SrCO$_3$ (99.994\% pure, Alfa Aesar) and Fe$_2$O$_3$ (99.998\% pure, Alfa Aesar) powders in air at 1050--1100$^\circ$C for at least 50 hours, with intermediate grinding.  Measurements of mass loss indicated that the reaction was nearly complete within the first 24 hours.  The resulting powder was packed into latex sleeves and pressed into rods under 125~MPa of hydrostatic pressure, then sintered in flowing oxygen at 1300$^\circ$C for 24 hours and cooled to room temperature.  These sintering conditions are similar to the conditions the rod encounters in the image furnace, and help prevent the rods from cracking apart during growth.  Sintered rods were approximately $\sim$9~mm in diameter and typically 10--15~cm long.

\begin{figure}[htb]
\includegraphics[width=\columnwidth]{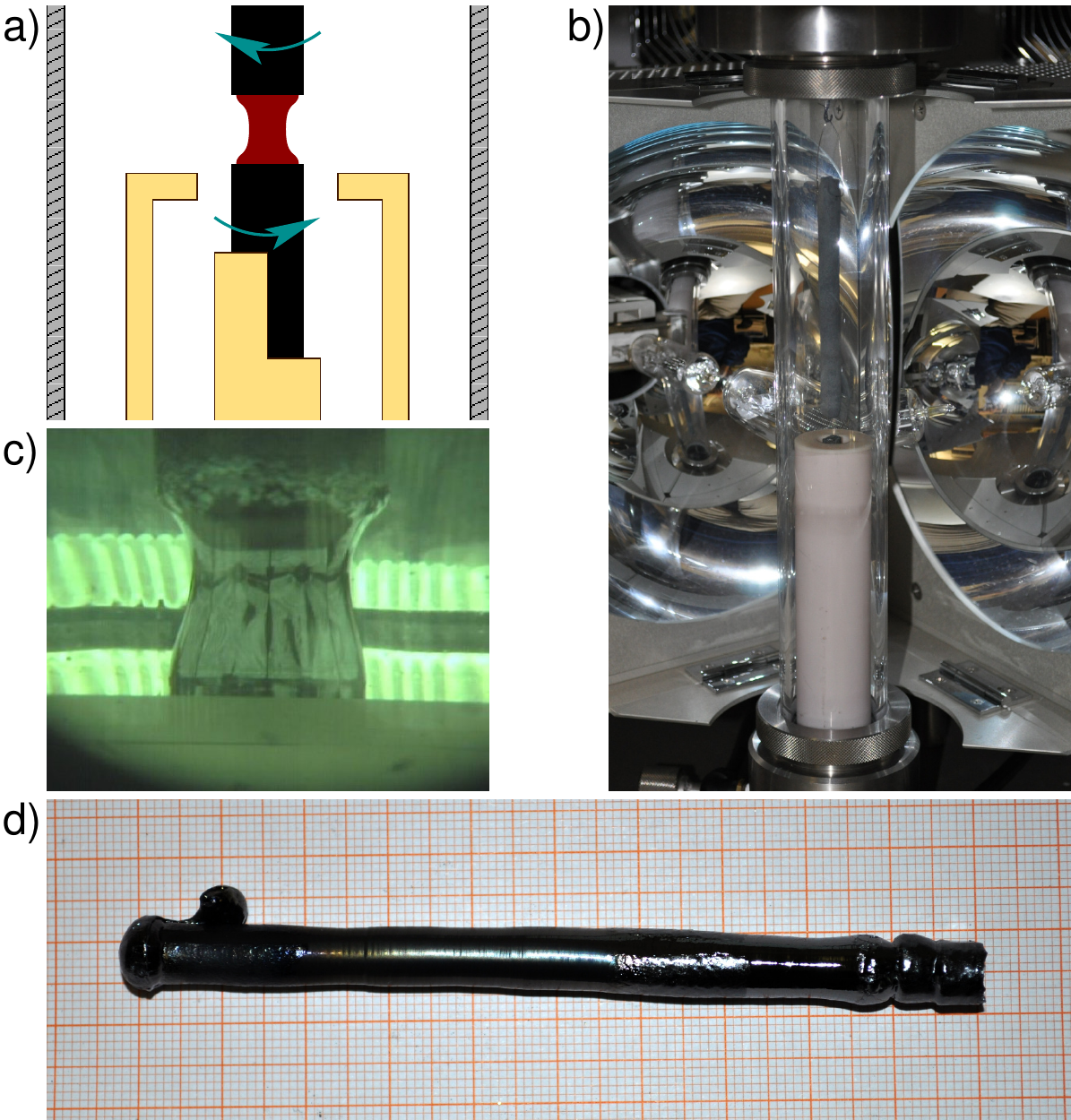}
\caption{\label{fig:shield}Floating zone growth with enhanced gradient.  a) Schematic diagram of image furnace growth setup, showing the feed and seed rods, molten zone, quartz tube, and the alumina shield used to increase the temperature gradient after growth.  b) Photograph of actual image furnace setup, with shield in place.  c)  Molten zone during growth.  d)  An example of an as-grown \SFO\ crystal grown using this method.}
\end{figure}

Single crystals of \SFO\ 8~mm in diameter and up to 100~mm in length were grown from these polycrystalline rods by the floating zone technique under $\sim$2.5~atm of pressure, flowing at 100~sccm, in a Crystal Systems FZ-T-10000-H-III-VPR 4-mirror image furnace with $4\times 1000$~W lamps.  A slow growth rate of about 2~mm/h was used for low mosaicity~\cite{Maljuk2003}, and the seed and feed rods were counter-rotated at 12--14~rpm.  Polycrystalline seed rods were typically used, due to concerns over cracking of single-crystalline seeds, and the grown crystals became single-domain within 1~cm.  To increase the temperature gradient immediately below the growth front and prevent the crystal from shattering, an alumina shield was devised (see Fig.~\ref{fig:shield}).  While strong temperature gradients would ordinarily make cracking a more serious problem, the shield is intended to accelerate the structural phase transitions, nearly quenching the crystal, leading to smaller twin domains and smaller, less well-connected cracks.  The resulting single-crystalline rods were found to be marginally more robust when grown under oxygen partial pressures of about 0.8~atm in argon, intended to avoid the Brownmillerite phase, just miss the orthorhombic-tetragonal phase separation region, and intersect the tetragonal-cubic phase separation region at the lowest possible temperature (see Ref.~\cite{Mizusaki1992}).  The growth direction was 5--15$^\circ$ from the cubic (100) direction, while facets could be either (100) or (110) faces, depending on the growth conditions.  The otherwise counterintuitive result that a greatly increased temperature gradient prevents shattering confirms that the cracking is attributable to the phase transitions and not thermal stress.

\begin{figure}[htb]
\includegraphics[width=\columnwidth]{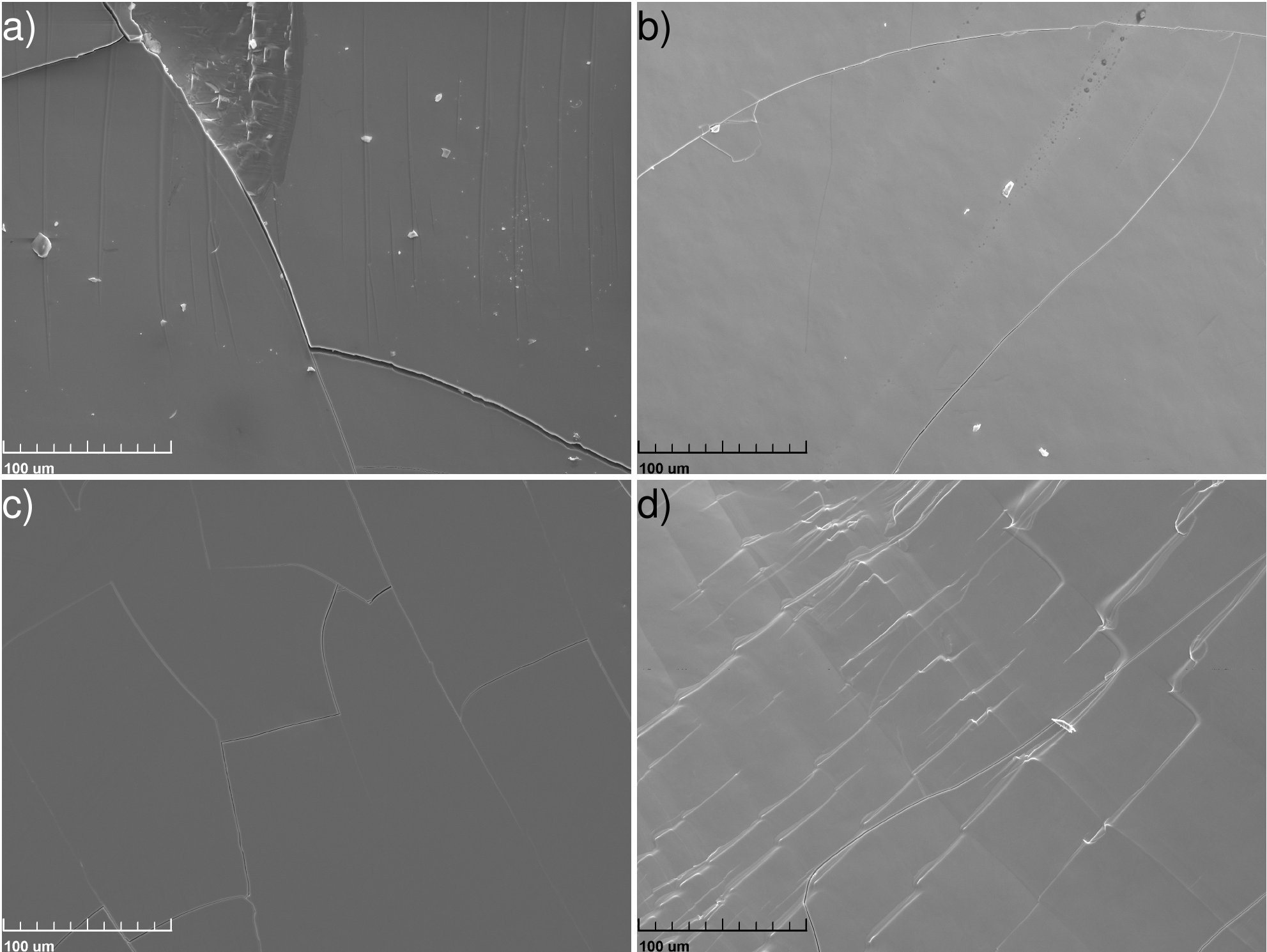}
\caption{\label{fig:SEM}SEM analysis of cracking:  a)  Crystal grown without a shield present, in 2.5~atm~O$_2$.  b)  and c)  Crystals grown with a shield present in 2.5~atm~O$_2$ --- density of cracks can vary substantially.  d)  Crystal grown with a shield present, in a partial pressure of 0.8~atm~O$_2$ in Ar.}
\end{figure}

Single-crystalline rods grown in this manner were sufficiently robust that they could be manipulated, cut with a diamond saw, annealed (including a quench into water), and clamped into sample holders without breaking.  Aside from qualitative observations of how robust the crystals were, the cracking in several as-grown crystals was characterized by scanning electron microscopy (SEM) in a VEGA TESCAN TS5130MM device, using an excitation voltage of 20~kV and a spot size of 32--34~nm.  Flat, as-broken surfaces from midway along the crystal rod were examined; unbroken samples were cracked apart by pinching with wire cutters as necessary.  Representative SEM images from several samples are compared in Fig.~\ref{fig:SEM}.  While it is difficult to extrapolate to the entire crystal from a few surfaces, the cracks were generally narrower when the shield was used;  no systematic trend was observed in the density of cracks.  Consequently, while crystals grown in this manner are suitable for neutron scattering and similar techniques requiring large sample masses, they do not constitute a significant advancement for fuel cell applications.

\section{Oxygen Annealing}

The crystals were annealed in alumina boats under controlled oxygen partial pressures and temperatures to ensure homogeneous, well-defined oxygen contents, based on the phase diagrams in Ref.~\cite{Mizusaki1992}.  The two dopings reported here, SrFeO$_{2.75}$ and SrFeO$_{2.81}$, were prepared under 0.002~atm O$_2$ in Ar at 449.2$^\circ$C, and at 475.0$^\circ$C in 1~atm O$_2$, respectively, for 7--10 days.  Since \SFO\ was observed to be insensitive to water, the crystals were quenched to deionized water at the conclusion of each anneal.  This occasionally caused slight chipping to the surfaces of larger pieces of crystal, but normally caused no obvious damage.  All characterization work reported here was performed on annealed crystals, with the exception of electron microscopy --- transport, magnetic and structural properties can depend on dopant homogeneity, so techniques sensitive to these require annealed crystals.  The oxygen contents produced by each anneal were verified by thermogravimetric (TG) analysis in a Netzsch STA-449C DTA/TG apparatus, by monitoring the mass loss while heating a sample in flowing argon (oxygen partial pressure $\sim10^{-7}-10^{-8}$~atm) to 1420$^\circ$C, under which conditions the oxygen content should be very close to SrFeO$_{2.50}$ and the time required for equilibration is very short.  A crystal intended to be SrFeO$_{2.75}$ had a measured oxygen content of 2.77, while the oxygen content of a sample with the nominal composition SrFeO$_{2.81}$ was found to be 2.82, well within the TG technique's estimated uncertainty of 0.02 of an oxygen atom per formula unit.  Given that the TG results confirm successful production of the intended oxygen contents, samples are referred to by their nominal stoichiometry.  

\begin{figure}[htb]
\includegraphics[width=\columnwidth,clip]{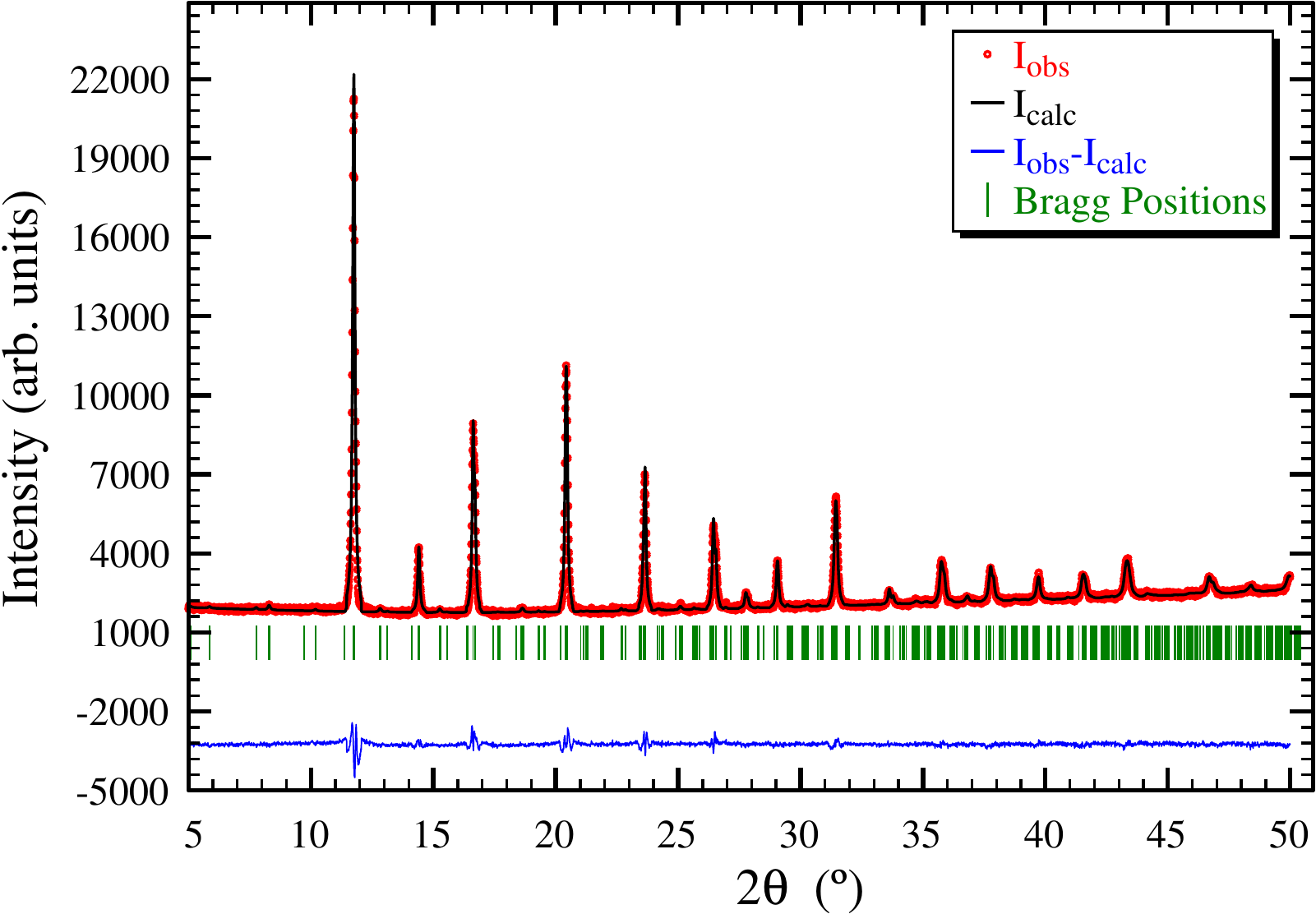}
\caption{\label{fig:powderXRD}Verification of phase purity:  Powder X-ray diffractogram on ground-up crystals of SrFeO$_{2.75}$, fit using the orthorhombic unit cell expected for this doping~\cite{Hodges2000}.  All peaks can be indexed, indicating phase purity.}
\end{figure}

\section{Characterization}

Phase purity was checked by room temperature x-ray powder diffraction on 0.5~mm-diameter glass capillaries containing powder ground from pieces of annealed crystals.  Data were collected from 2$^\circ$ to 50$^\circ$ in 2$\theta$ in steps of 0.01$^\circ$, with a Stoe Stadi-P diffractometer in Debye-Scherrer geometry, using silver K$\alpha$1 radiation and a Ge (111) monochromator, and detected using a linear position-sensitive detector with an opening of $\sim12^\circ$ in 2$\theta$.  A diffractogram for SrFeO$_{2.75}$ is shown in Fig.~\ref{fig:powderXRD}, fit assuming the published orthorhombic unit cell~\cite{Hodges2000} --- the dominant phase in a SrFeO$_{2.75}$ sample should be Sr$_4$Fe$_4$O$_{11}$, which has been reported as orthorhombic~\cite{Takeda1986}.  All peaks can be indexed, with none arising from impurity phases, demonstrating the sample's phase purity.  X-ray and neutron powder structure refinements at a variety of oxygen contents have been previously published~\cite{Takeda1986,Takano1988b,Hodges2000} and need not be repeated here;  a more thorough single-crystal diffraction study at several dopings has commenced~\cite{Reehuis2012}, and full structure refinements will be performed on the crystals grown in this study as the diffraction study progresses.  Energy-dispersive x-ray (EDX) composition mapping was also used to verify the phase purity and homogeneity of several unannealed crystals --- no inhomogeneity was observed.

\begin{figure}[htb]
\includegraphics[width=\columnwidth]{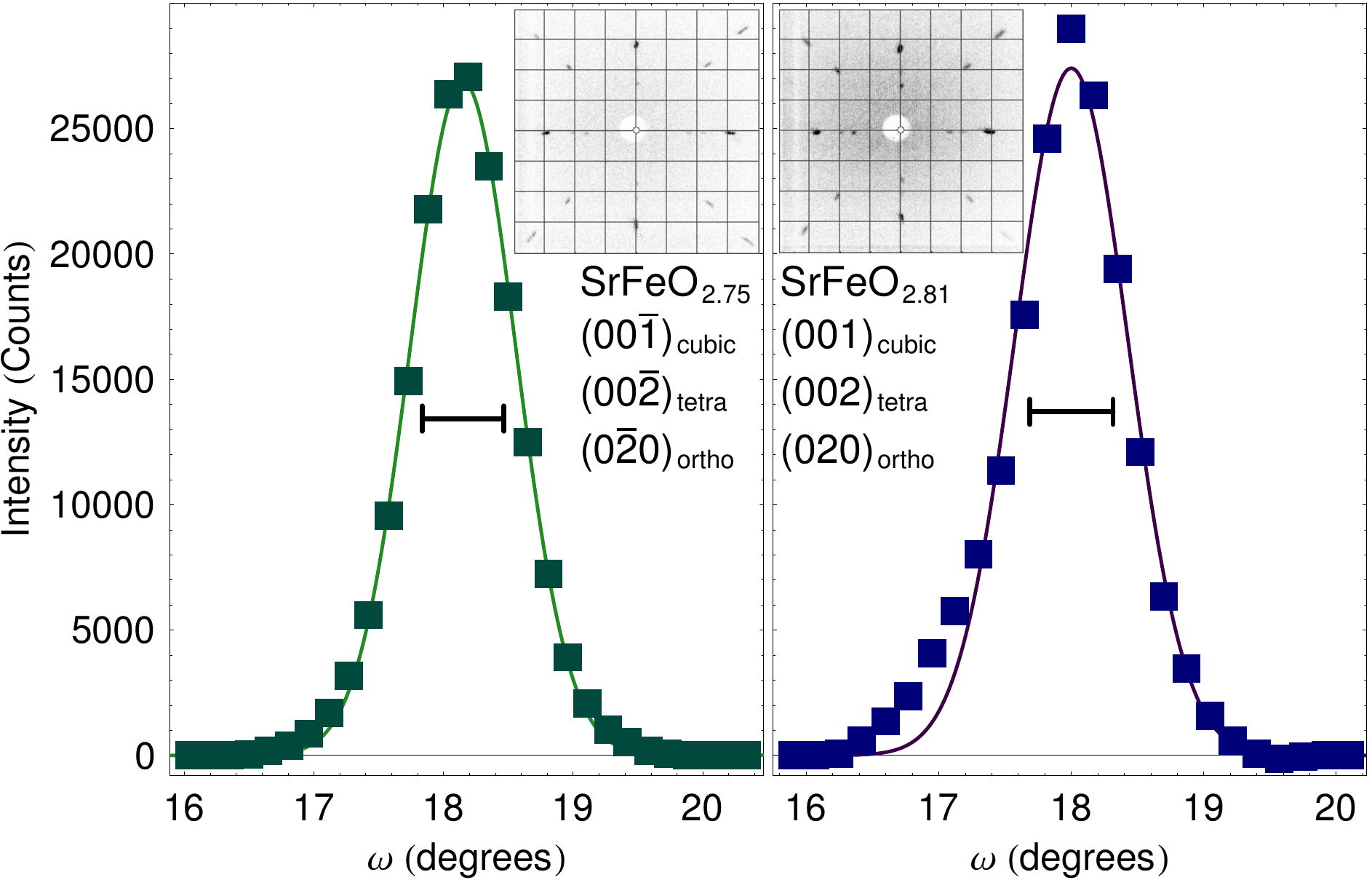}
\caption{\label{fig:rocking}Crystallinity:  Neutron rocking curves for $\sim$0.3~cm$^3$ crystals of SrFeO$_{2.75}$ (left), (00\=1)$_{cubic}$ = (00\=2)$_{tetra}$ = (0\=20)$_{ortho}$ reflection fit to a Gaussian distribution indicating a full width at half maximum (FWHM) of $0.957\pm0.008^\circ$; and SrFeO$_{2.81}$ (right), (001)$_{cubic}$ = (002)$_{tetra}$ = (020)$_{ortho}$ reflection, FWHM $1.00\pm0.03^\circ$.  The instrument resolution is indicated.  Insets to these panels show x-ray Laue images along (100)$_{cubic}$ for SrFeO$_{2.75}$ and SrFeO$_{2.81}$ crystals, respectively.}
\end{figure}

The crystallinity of several large pieces of crystal was checked using neutron diffraction on the four-circle diffractometer E5 at the Helmholtz-Zentrum Berlin's BER~II reactor, using a pyrolytic graphite (002) monochromator to select a wavelength of 2.36~\AA.  A SrFeO$_{2.75}$ (00\=1)$_{cubic}$ = (00\=2)$_{tetra}$ = (0\=20)$_{ortho}$ and a SrFeO$_{2.81}$ (001)$_{cubic}$ = (002)$_{tetra}$ = (020)$_{ortho}$ rocking curve is shown in Fig.~\ref{fig:rocking}.  The former crystal was $\sim$8~mm diameter $\times$ 7~mm long and the latter $\sim$6~mm diameter $\times$ 5~mm.  The full widths at half-maximum (FWHM) for these peaks from Gaussian fits to the data were $0.957\pm0.008^\circ$ and $1.00\pm0.03^\circ$, respectively, and the instrumental resolution of 0.63$^\circ$ at this angle is indicated.  Broadening is likely attributable to twinning, which would be expected at this doping, although a shoulder indicates that the SrFeO$_{2.81}$ crystal includes a small secondary grain.  Twinning makes it impossible to distinguish among peaks in the oxygen-ordered phases that would be classified as (100)-type in the cubic setting:  (220) and (002) in the tetragonal setting and (201) and (020) in the orthorhombic.  X-ray Laue photos along the (100) axes of a SrFeO$_{2.75}$ and a SrFeO$_{2.81}$ crystal are shown in insets.  The quality of the rocking curves and Laue photos indicate good crystallinity.

\section{Magnetic and Transport Properties}

\begin{figure}[htb]
\includegraphics[width=\columnwidth]{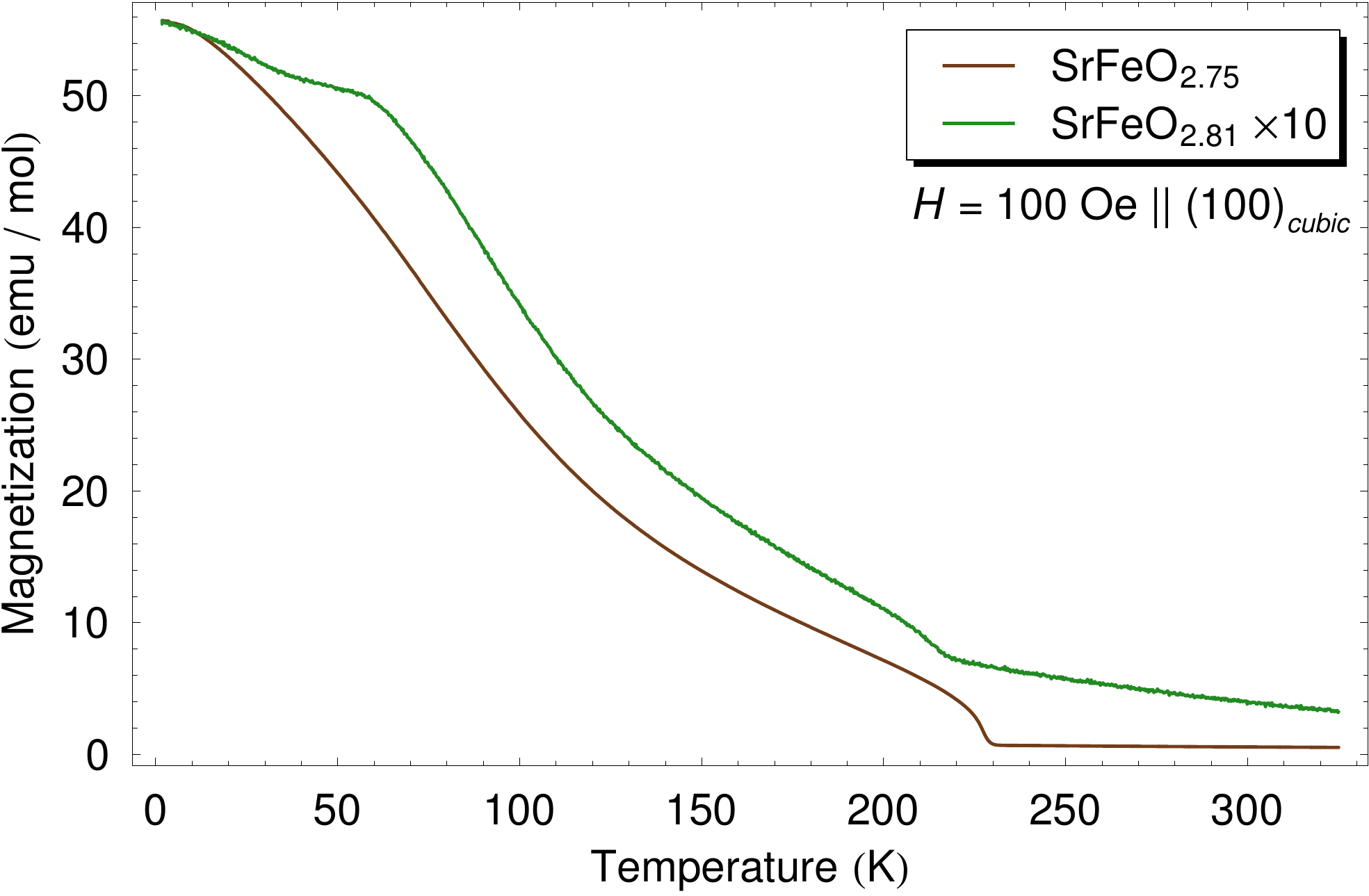}
\caption{\label{fig:squid_both}Magnetic response:  Magnetization of samples of quenched SrFeO$_{2.75}$ and SrFeO$_{2.81}$, cooled in an applied field of 100~Oe $\parallel (100)_{cubic}$.}
\end{figure}

The field-cooled magnetic response of samples with the two oxygen contents was measured on cooling using a SQUID magnetometer (Quantum Design MPMS-VSM) in an applied field of 100~Oe along the cubic $a$~axis.  Crystals were mounted on a long quartz slab with teflon tape and centred in the machine at 300~K in a field of 2000~Oe, then the field was oscillated down to 100~Oe to avoid any field-training.  The results are presented in Fig.~\ref{fig:squid_both}.  Both traces closely resemble that reported for a SrFeO$_{2.75}$ powder sample in the same applied field in Ref.~\cite{Schmidt2003}, with SrFeO$_{2.75}$ and SrFeO$_{2.81}$ exhibiting jumps in their magnetization at 230~K and 218~K, respectively, compared to the earlier report of 232~K.  An additional small hump is visible near 60~K in the SrFeO$_{2.81}$ data (in stronger applied fields, this is more pronounced).  Two forms of magnetic order with transition temperatures of 60~K and 65~K have recently been reported in this system~\cite{Reehuis2012};  one of these may be responsible.  Note that the two dopings' vertical scales differ by an order of magnitude.

\begin{figure}[htb]
\includegraphics[width=\columnwidth]{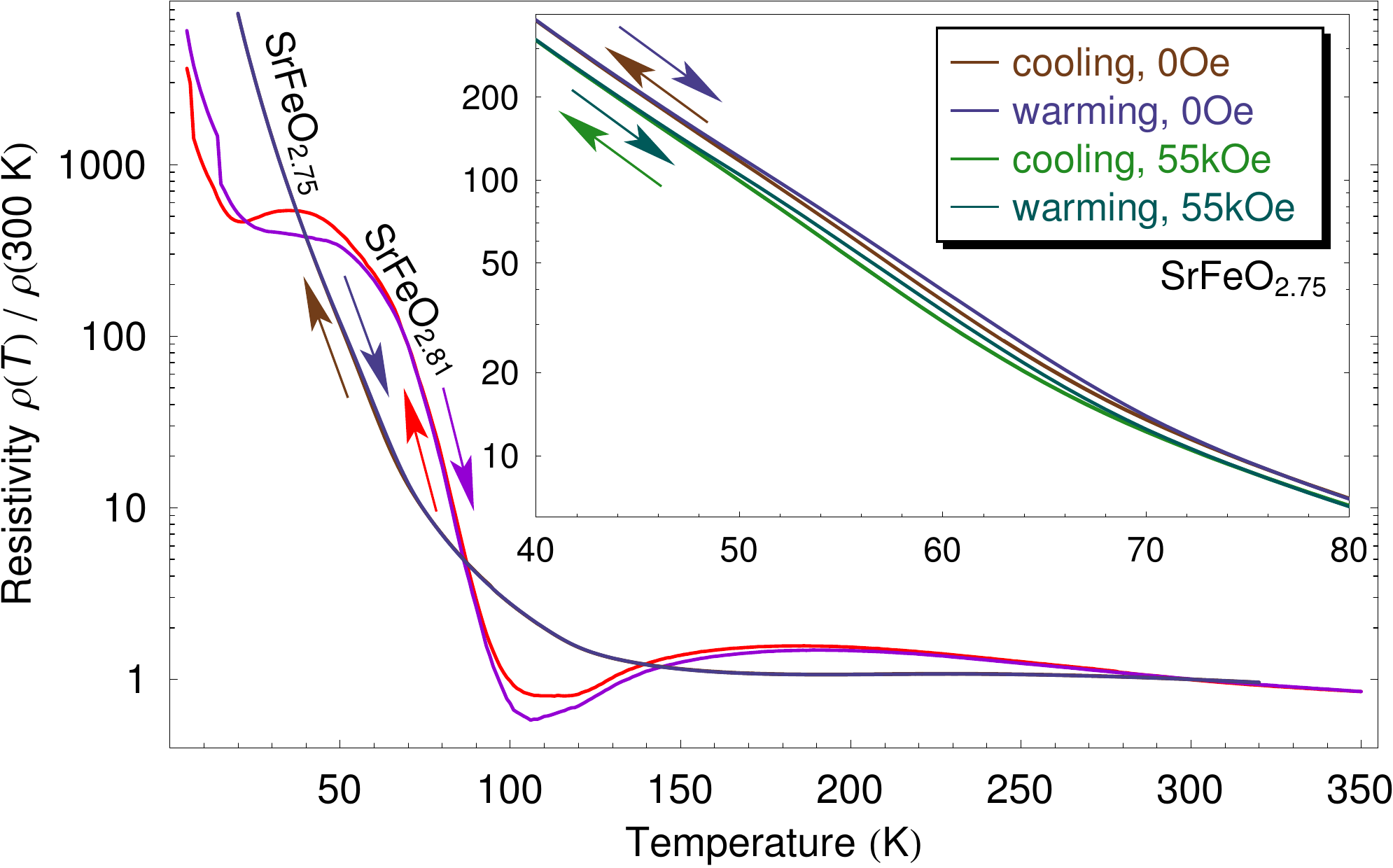}
\caption{\label{fig:rho}Electrical transport:  Zero-field resisitivity of single crystals of SrFeO$_{2.75}$ and SrFeO$_{2.81}$, each normalized to its value at 300~K.  Note that this normalization prevents a comparison of the two samples' absolute resistivities.  The resistivity of each crystal increases strongly below $\sim$175~K (SrFeO$_{2.75}$) or $\sim$105~K (SrFeO$_{2.81}$).  SrFeO$_{2.75}$ exhibits very weak hysteresis between about 50 and 70~K, which is insensitive to applied field (inset), possibly due to minor contamination by a second magnetic phase.}
\end{figure}

The resistivity of a roughly $0.8\times0.6\times0.3$~mm$^3$ single crystal of SrFeO$_{2.75}$ was measured using a Quantum Design MPMS SQUID magnetometer for temperature and field control, and a Conductus LTC-20 low temperature controller as a resistance bridge for data collection;  a SrFeO$_{2.81}$ crystal of similar dimension was measured in a Quantum Design PPMS.  Gold leads were attached to corners of the crystal using silver epoxy, which covered the corner from top to bottom to help ensure uniform two-dimensional transport.  The silver epoxy was allowed to cure for eight hours at 250$^\circ$C in air --- it is important to note that, while thermogravimetric analysis and other tests indicate that this is not a high enough temperature or long enough time for any significant quantity of oxygen to enter or leave the sample, it is possible that oxygen atoms may be able to hop from site to site at this temperature, possibly altering the oxygen order, increasing twin domain size, or relieving strain, so on a microscopic level the resistivity samples may be different from other samples characterized.  It is also important to note that microcracking complicates the conversion of measured transport data into absolute quantities such as resistivity, since the actual path length and cross-sectional area of the measurement cannot be known.  It has previously been observed that reducing the oxygen stoichiometry from 3.00 rapidly introduces a low-temperature transition to semiconducting or insulating behaviour~\cite{MacChesney1965,Oda1977,Lebon2004,Zhao2004,Adler2006}, and Fig.~\ref{fig:rho} indeed shows the samples' resistivity diverging at low temperatures.  The resistivity of SrFeO$_{2.75}$ exhibits very weak but reproducible hysteresis between about 50 and 70~K (see Fig.~\ref{fig:rho} inset), possibly due to a very small volume fraction of a second magnetic phase introduced by imperfect annealing or when curing the contacts.  The hysteresis in SrFeO$_{2.75}$ is essentially unchanged by the application of a magnetic field, as can be seen in the Fig.~\ref{fig:rho} inset, which compares the resistive response in this region in applied fields of 0~Oe and 55~kOe.  A hysteretic transition around this temperature has been observed previously at slightly higher dopings~\cite{Adler2006} but not identified, and two forms of magnetic order with onset temperatures of 60 and 65~K have been observed but not linked to any oxygen-ordered phase~\cite{Reehuis2012}.  Full identification of this phase transition is the subject of ongoing research.  The SrFeO$_{2.81}$ crystal's resistivity is far more complicated --- this doping range is prone to mixing of oxygen-ordered phases with very different transport properties~\cite{Adler2006}.

\section{Summary}

In conclusion, we have succeeded in growing complete single-crystalline rods of \SFO\ 8~mm in diameter and up to 10~cm in length, avoiding the shattering problems typically encountered in this system.  The samples are still cracked, but in such a way that they hold together.  An annealing process is described, and its results characterized by magnetic and transport measurements.  The availability of high-quality, well-annealed crystals of the size reported here will enable future studies requiring large sample masses, such as the investigation by neutron scattering of their magnetic structure and magnetic excitations.  

\section*{Acknowledgements}

The authors thank Y.\ Liu, A.\ Maljuk, and C.\ T.\ Lin for helpful advice, C.\ Stefani and C.\ Busch for assistance in sample characterization, and the staff of the MPI-FKF Crystal Growth service group.  The authors gratefully acknowledge the support of the MPI-UBC Quantum Materials Institute.

\bibliographystyle{apsrev4-1}
\bibliography{ferrate}

\end{document}